\documentclass[aps,nofootinbib,prd,superscriptaddress,tightenlines,notitlepage,onecolumn,showpacs]{revtex4-2}

\usepackage{graphicx}
\usepackage{dcolumn}
\usepackage{bm}
\usepackage[colorlinks=true, linkcolor=blue, citecolor=blue]{hyperref}

\begin{document}
\newcommand{\apjl}{Astrophys. J. Lett.}
\newcommand{\apjs}{Astrophys. J. Suppl. Ser.}
\newcommand{\apss}{Astrophysics and Space Science}
\newcommand{\aap}{Astron. \& Astrophys.}
\newcommand{\aj}{Astron. J.}
\newcommand{\araa}{Ann. Rev. Astron. Astrophys.} 
\newcommand{\mnras}{Mon. Not. R. Astron. Soc.}
\newcommand{\jcap}{JCAP}
\newcommand{\pasj}{PASJ}
\newcommand{\pasa}{Pub. Astro. Soc. Aust.}
\newcommand{\pasp}{Pub. Astro. Soc. Pacific}
\newcommand{\physrep}{Physics Reports}
\newcommand{\ssr}{Space Sci. Rev.}
\newcommand{\aapr}{Astron. Astrophys. Rev.}
\newcommand{\jgr}{J. Geophys. Res.}
\newcommand{\cjaa}{Chin. J. Astron. Astrophys.}
\newcommand{\lsi}{LS I +61$^{\circ}$ 303}
\preprint{APS/123-QED}

\title{First detection of ultra-high energy emission from gamma-ray binary \lsi}
 
\author{Zhen Cao}
\affiliation{Key Laboratory of Particle Astrophysics \& Experimental Physics Division \& Computing Center, Institute of High Energy Physics, Chinese Academy of Sciences, 100049 Beijing, China}
\affiliation{University of Chinese Academy of Sciences, 100049 Beijing, China}
\affiliation{TIANFU Cosmic Ray Research Center, Chengdu, Sichuan,  China}
 
\author{F. Aharonian}
\affiliation{TIANFU Cosmic Ray Research Center, Chengdu, Sichuan,  China}
\affiliation{University of Science and Technology of China, 230026 Hefei, Anhui, China}
\affiliation{Yerevan State University, 1 Alek Manukyan Street, Yerevan 0025, Armenia}
\affiliation{Max-Planck-Institut for Nuclear Physics, P.O. Box 103980, 69029  Heidelberg, Germany}
 
\author{Y.X. Bai}
\affiliation{Key Laboratory of Particle Astrophysics \& Experimental Physics Division \& Computing Center, Institute of High Energy Physics, Chinese Academy of Sciences, 100049 Beijing, China}
\affiliation{TIANFU Cosmic Ray Research Center, Chengdu, Sichuan,  China}
 
\author{Y.W. Bao}
\affiliation{Tsung-Dao Lee Institute \& School of Physics and Astronomy, Shanghai Jiao Tong University, 200240 Shanghai, China}
 
\author{D. Bastieri}
\affiliation{Center for Astrophysics, Guangzhou University, 510006 Guangzhou, Guangdong, China}
 
\author{X.J. Bi}
\affiliation{Key Laboratory of Particle Astrophysics \& Experimental Physics Division \& Computing Center, Institute of High Energy Physics, Chinese Academy of Sciences, 100049 Beijing, China}
\affiliation{University of Chinese Academy of Sciences, 100049 Beijing, China}
\affiliation{TIANFU Cosmic Ray Research Center, Chengdu, Sichuan,  China}
 
\author{Y.J. Bi}
\affiliation{Key Laboratory of Particle Astrophysics \& Experimental Physics Division \& Computing Center, Institute of High Energy Physics, Chinese Academy of Sciences, 100049 Beijing, China}
\affiliation{TIANFU Cosmic Ray Research Center, Chengdu, Sichuan,  China}
 
\author{W. Bian}
\affiliation{Tsung-Dao Lee Institute \& School of Physics and Astronomy, Shanghai Jiao Tong University, 200240 Shanghai, China}
 
\author{A.V. Bukevich}
\affiliation{Institute for Nuclear Research of Russian Academy of Sciences, 117312 Moscow, Russia}
 
\author{C.M. Cai}
\affiliation{School of Physical Science and Technology \&  School of Information Science and Technology, Southwest Jiaotong University, 610031 Chengdu, Sichuan, China}
 
% ...
\collaboration{The LHAASO Collaboration}

%
%
%
%

\email[E-mail: ]{dongxuqiang@ihep.ac.cn; gmxiang@ihep.ac.cn; licong@ihep.ac.cn; zjn@shao.ac.cn; hhh@ihep.ac.cn}

\begin{abstract}
We report the first detection of gamma-ray emission up to ultra-high-energy (UHE; $>$100 TeV) emission from the prototypical gamma-ray binary system \lsi\ using data from the Large High Altitude Air Shower Observatory (LHAASO). It is detected with significances of 9.2$\sigma$ in WCDA (1.4--30.5 TeV) and 6.2$\sigma$ in KM2A (25--267 TeV); in KM2A alone we identify 16 photon-like events above 100 TeV against an estimated 5.1 background events, corresponding to a 3.8$\sigma$ detection. These results provide compelling evidence of extreme particle acceleration in \lsi. Furthermore, we observe orbital modulation at 3.9$\sigma$ confidence level, between 25 and 100 TeV, with a hint that the orbital modulation is energy-dependent. These features can be understood in a composite scenario in which leptonic and hadronic processes jointly contribute. 
\end{abstract}


\maketitle


\section{\label{sec1}Introduction}
Binary systems are prominent sources in the high-energy sky above a few keV, with X-ray binaries being the dominant population \cite{Grimm_2002}. However, only a small number of these systems have been detected at more energetic gamma-ray bands \cite{Dubus_2013,2019A&A...631A.177C}. Among them, a distinct class known as gamma-ray binaries has been identified, characterized by persistent gamma-ray emission and spectral energy distributions ($\nu F_{\nu}$) that peak in the gamma-ray range \cite{Dubus_2013,2020mbhe.confE..45C}.
All known gamma-ray binaries consist of a compact object in orbit around a massive, young O- or Be-type star, with clear orbital modulation observed across multiple wavelengths \cite{Dubus_2013}. These observed modulations indicate that the gamma-ray emission is likely powered by intra-binary interactions --- such as collisions between a relativistic outflow from the compact object and the dense stellar wind or radiation field of the companion star. 
TeV emission was detected from these compact systems, but it is unclear about the maximal energy reachable by binary systems and whether they are capable of accelerating protons to PeV energies. Probing these systems at ultra-high-energy (UHE; $>$100 TeV) range opens a unique window to study extreme particle acceleration, relativistic outflows, and the accretion–ejection processes operating in compact binary environments.

\lsi\ is a prototypical gamma-ray binary system composed of a massive Be-type star and a compact object of uncertain nature \cite{1994A&A...288..519P, Gregory_2002}. Recent report of transient radio pulsations from the direction of \lsi\ strongly suggests that the compact object is a rotating neutron star \cite{Weng:2022zlg}. Located at a distance of $\sim$2 kiloparsecs \cite{1998MNRAS.297L...5S}, \lsi\ exhibits a well-established orbital period of 26.496(8) days \cite{1984ApJ...283..273T}, confirmed across optical \cite{1989MNRAS.239..733M}, X-ray \cite{2006ATel..940....1L,2014ApJ...785L..19L}, and gamma-ray \cite{Fermi-LAT:2009ldx,2024ApJ...972...80Z} bands. In 2006, the MAGIC telescope detected variable TeV gamma-ray emission modulated by the orbital period, with the short variability time scale suggesting a compact emission origin \cite{MAGIC:2006pue}. This orbital modulation was subsequently confirmed by VERITAS \cite{Acciari:2008hg}. Both instruments detected enhanced TeV emission near apastron.
Besides the well-established orbital modulation, \lsi\ is also known to exhibit significant superorbital variability on timescales of $\sim 4.5$ years \cite{MAGIC:2016oil}, as well as stochastic TeV flares \cite{VERITAS:2016gnm}.

The Large High Altitude Air Shower Observatory (LHAASO) is a large hybrid observatory designed to explore the gamma-ray sky from $\sim 100$ GeV to several PeV \cite{LHAASO:2019qtb}. It consists of three major detector arrays: the Kilometer Square Array (KM2A), the Water Cherenkov Detector Array (WCDA) and the Wide Field-of-View Cherenkov Telescope Array (WFCTA) \cite{LHAASO:2021ozi, LHAASO:2024zug, Aharonian:2020iou}. In this paper, we present LHAASO observations of \lsi, spanning roughly 1 TeV to several hundred TeV. These results provide unprecedented access to the UHE regime and offer new constraints on the nature of particle acceleration in gamma-ray binaries.

\section{Methods}\label{sec3}

Data from both WCDA and KM2A are used and analyzed independently in this work. The WCDA data are collected with the full array from March 2021 to July 2024, using the number of triggered detectors ($N_{\rm{hit}}$) as a proxy for the energy estimate. The KM2A data used here were collected from December 2019 to July 2024, covering three phases of the array. For KM2A, the primary energy is reconstructed based on the particle density at a fixed distance of 50 m from the shower axis.
The data are binned with a resolution of five bins per energy decade. The residual cosmic-ray background after $\gamma$/hadron separation is estimated using the `direct integration method' for both arrays \cite{Fleysher:2003nh}. (Further details regarding the LHAASO detector and event reconstruction are provided in the Supplementary Material and previous performance studies \cite{LHAASO:2021ozi, Aharonian:2020iou}.)

The galactic diffuse emission(GDE), produced by the interaction of the cosmic ray `sea' with interstellar gas, may also influence the results \cite{Cygnus_Bubble}. The GDE is assumed to follow the spatial distribution of the Planck dust optical depth map, while its spectrum is modeled by a power-law function. In our joint likelihood fit, both the normalization and the spectral index of the GDE are treated as free parameters. A circular region with a radius of $4^{\circ}$ around \lsi\ is chosen as the Region of Interest (ROI). A 3D likelihood fitting process is employed in this analysis, in which the morphological and spectral parameters are fitted simultaneously \cite{LHAASO:2023rpg}. Likelihood ratio tests are performed for each potential source in the ROI. A source is added to the model if the improvement in the Test Statistic (TS) exceeds 25 (see Supplementary Material for details).

The sky map in the celestial coordinates is divided into $0.1^{\circ} \times 0.1^{\circ}$ pixels. The TS value at each pixel is calculated using the same likelihood fitting method described above. Specifically, the null hypothesis consists of the background-only model. In the alternative hypothesis, in addition to the background, a point source with its spectral index fixed to the best-fit value is added at the position of each pixel. Since the normalization is the only free parameter, the significance is estimated as $\sqrt{TS}$ according to Wilks' theorem.

To maintain consistent detector performance, only data taken with the full array from July 2021 to July 2024 are used for the orbital-variability analysis. The orbital phase of each event is calculated as $\phi = ((T-T_{0}) \bmod P)/P$, where $T_{0} = 43366.275$ MJD is the zero-phase epoch \cite{1982ApJ...255..210T}, $P=26.496$ d is the orbital period \cite{1984ApJ...283..273T}, and $\bmod$ denotes the modulo operator. We adopt the widely accepted periastron phase $\phi_{\rm peri} = 0.23$ \cite{2005MNRAS.360.1105C,MAGIC:2016oil}, although alternative orbital solutions yield a slightly later phase ($\phi_{\rm peri} = 0.275$; \cite{2009ApJ...698..514A}).
The data are divided into ten phase bins per orbit. A likelihood-ratio test is employed, where the null hypothesis assumes constant flux across all bins, whereas the alternative hypothesis allows the flux to vary independently among them. The flux normalization for each phase bin is fitted by assuming the average photon index derived from the full dataset. Under Wilks' theorem, the resulting TS follows a $\chi^2$ distribution with 9 degrees of freedoms.

\begin{figure}[h]
\centering
\includegraphics[width=0.45\textwidth]{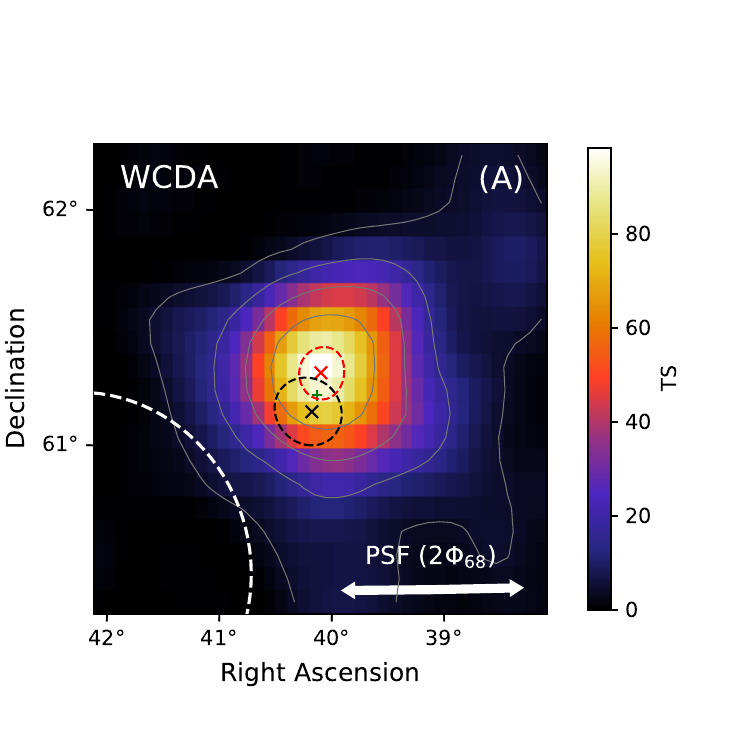}
\includegraphics[width=0.45\textwidth]{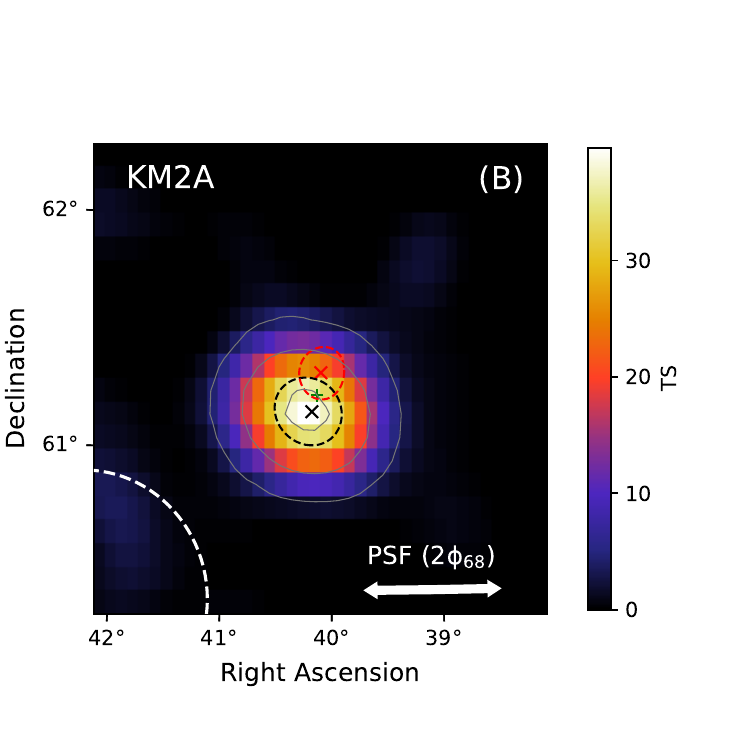}
\caption{
Test-statistic maps of the region around \lsi\ observed by WCDA {\bf (A)} and KM2A {\bf (B)}, after subtracting the contributions from nearby sources and Galactic diffuse emission. The energy ranges are 1.5–30.5 TeV for WCDA and $>25$ TeV for KM2A, with a sky map pixel size of $0.1^\circ \times 0.1^\circ$. The gray solid lines represent the significance contours starting from $2\sigma$ with a step of $2\sigma$ (reaching $8\sigma$ for WCDA and $6\sigma$ for KM2A). Green crosses indicate the optical position of \lsi\ \cite{Gaia:2023fqm}. The red and black symbols represent the results for WCDA and KM2A, respectively: `x' marks the best-fit positions of \lsi, while dashed circles illustrate the 95\% confidence level positional uncertainties, incorporating both statistical fitting errors and systematic pointing uncertainties. White dashed lines indicate the 68\% containment regions (convolved with the PSF) of the extended source 1LHAASO J0249+6022. The PSF for each detector is marked by a double-headed arrow.}
\label{fig1}
\end{figure}

\begin{figure}[h]
\centering
\includegraphics[width=0.8\textwidth]{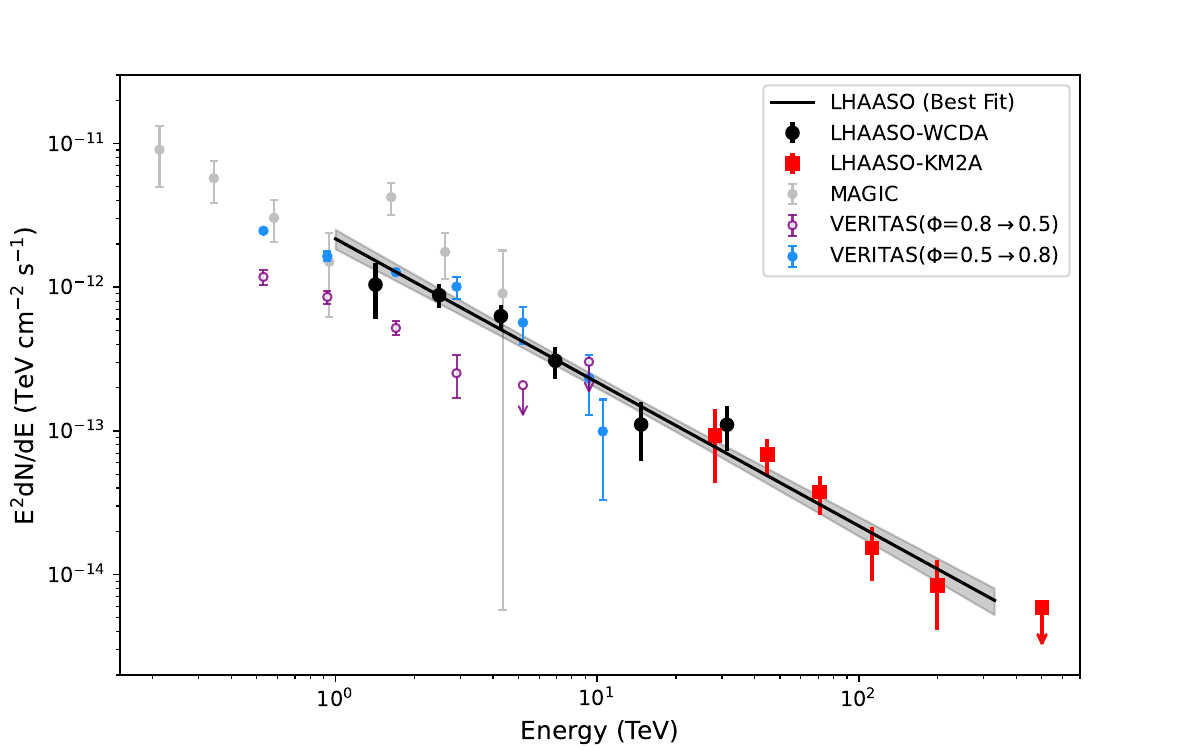}
\caption{Spectral energy distribution of \lsi\ observed by LHAASO. The measurements from WCDA and KM2A are marked with black dots and red squares, respectively. The solid line represents the best-fit result assuming a power-law distribution, while the butterfly plot shows the 1$\sigma$ statistical uncertainties. For comparison, measurements from MAGIC (marked with gray points; \cite{MAGIC:2006pue}) and VERITAS (marked with purple circles for orbital phases from 0.8 to 0.5, and blue points for orbital phases from 0.5 to 0.8; \cite{2017ICRC...35..712K}) are shown.}
\label{fig2}
\end{figure}

\begin{figure}[h]
\centering
\includegraphics[width=0.8\textwidth]{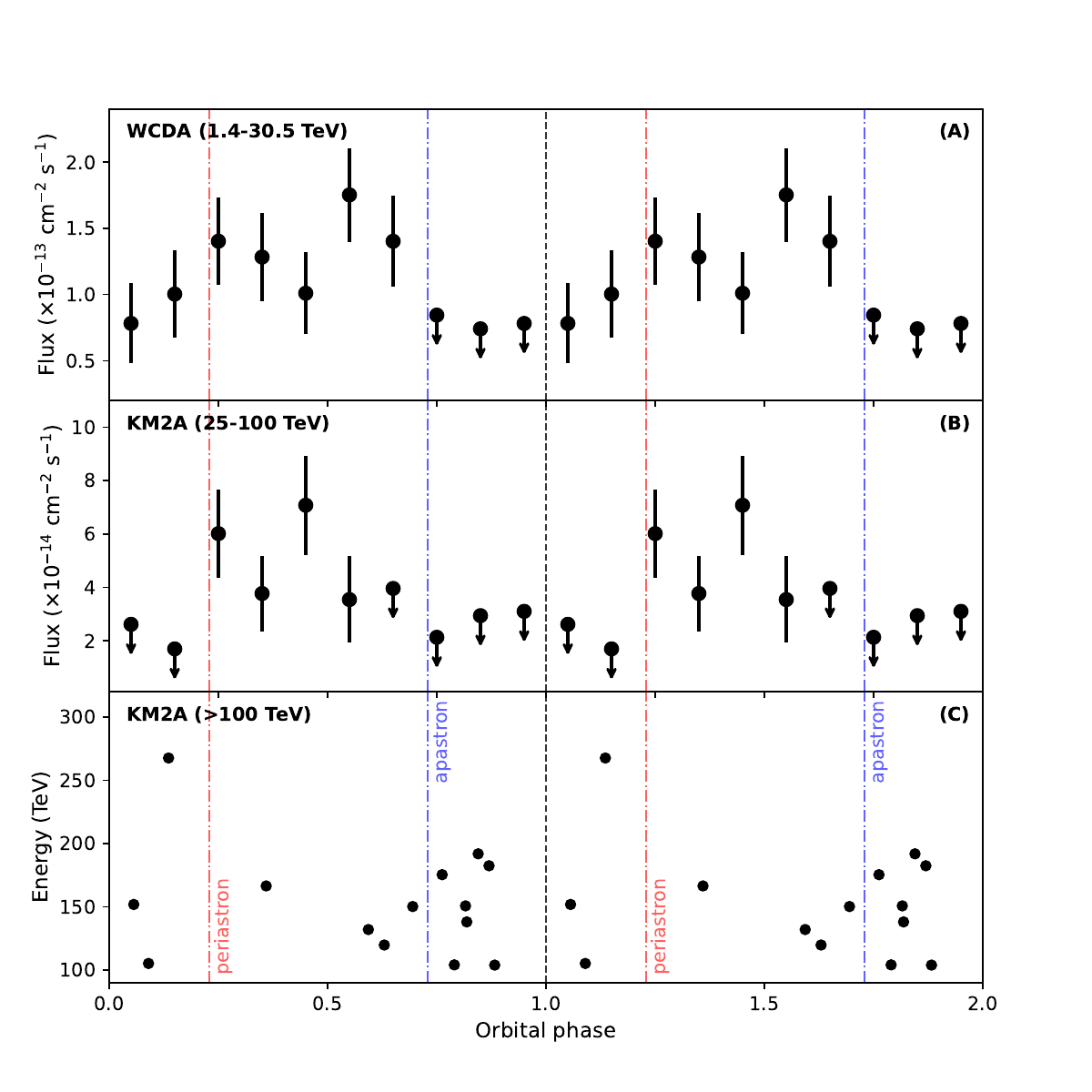}
\caption{Phase-resolved emission of \lsi\ across different energy bands. ({\bf A,B}) Gamma-ray flux measured by WCDA (1.5--30.5 TeV) and KM2A (25--100 TeV) as a function of the orbital phase. Downward arrows represent the 95\% confidence level upper limits. (C) Estimated energy of individual photon $>100$ TeV detected by KM2A versus their corresponding orbital phases. For clarity, two orbital cycles are shown. The red and blue dash-dotted vertical lines indicate the periastron and apastron phases, respectively, according to the orbital solution from \cite{2005MNRAS.360.1105C}.}
\label{fig3}
\end{figure}
\noindent

\section{Results}\label{sec11}
Fig.\ref{fig1} shows the significance maps around \lsi\ at two different energies. A total of two sources are resolved within the ROI: 1LHAASO J0249+6022 and \lsi. Specifically, \lsi\ is detected with a significance of 9.2$\sigma$ by WCDA and 6.2$\sigma$ by KM2A. At energies above 100 TeV, 16 photon-like events are detected from the direction of \lsi\ against an expected cosmic-ray background of 5.1 events, corresponding to a significance of 3.8$\sigma$ according to the Li \& Ma formula \cite{LiMa}. The total contribution from 1LHAASO J0249+6022 ($\sim 1.9^{\circ}$ away) and the Galactic diffuse emission is estimated to be only 0.12 events above 100 TeV. The highest-energy photon-like event, with an estimated energy of $267 \pm 61$ TeV, was detected at an angular distance of 0.09$^{\circ}$ from \lsi. This deviation is consistent with the detector's Point Spread Function (PSF), which has a 68\% containment radius of 0.17$^{\circ}$ at this energy. We estimate the energy uncertainty based on simulations of the instrument’s response.
The best-fit position determined by WCDA using a 2D Gaussian model is R.A. = $40.10^{\circ} \pm 0.07^{\circ}_{\rm stat}$ and Decl. = $61.33^{\circ} \pm 0.04^{\circ}_{\rm stat}$, while for KM2A it is R.A. = $40.18^{\circ} \pm 0.10^{\circ}_{\rm stat}$ and Decl. = $61.16^{\circ} \pm 0.05^{\circ}_{\rm stat}$. After calibration using events at comparable zenith angles (see Supplementary Material), the pointing accuracies are 0.03$^{\circ}$ for WCDA and 0.05$^{\circ}$ for KM2A. No significant spatial extension is detected; the 95\% confidence level upper limits on extension width are 0.22$^\circ$ for WCDA and 0.25$^\circ$ for KM2A.

The flux points from WCDA and KM2A are shown in Fig.~\ref{fig2}. The quoted errors are statistical only. Systematic uncertainties are assessed independently for WCDA and KM2A using the Crab Nebula as a standard candle, yielding $^{+8\%}_{-5\%}$ and $\pm 7\%$ for flux measurements, respectively. In addition, the use of different spatial models for 1LHAASO J0249+6022 results in a 6$\%$ uncertainty for flux measurement for WCDA. While for KM2A, this influence is negligible. The difference between the atmospheric model used in the simulation and the actual atmosphere dominates the systematic error for KM2A; this introduces an uncertainty of 0.02 in the spectral index. The combined flux points from 1 TeV to hundreds of TeV are well-described ($\chi^2/\text{dof} = 6.7/8$) by a single power-law model: $dN/dE = N_{0} (E/10\,\text{TeV})^{-\Gamma}$. The best-fit parameters are $N_{0} = (2.18 \pm 0.20) \times 10^{-15}\,\text{TeV}^{-1}\,\text{cm}^{-2}\,\text{s}^{-1}$ for the flux normalization and $\Gamma = 3.00 \pm 0.05$ for the photon index. The integral flux above 1 TeV, $F_{>1\,\text{TeV}} = 1.10 \times 10^{-12}\,\text{cm}^{-2}\,\text{s}^{-1}$, corresponds to $3.8\%$ of the Crab Nebula flux (Crab unit). Fig.~\ref{fig2} also includes spectra measured by VERITAS and MAGIC for comparison. Although the source exhibits clear orbital modulation and its orbitally-averaged flux shows slight variations on long timescales (i.e., super-orbital modulation; \cite{2012ApJ...744L..13L, MAGIC:2016oil}), the TeV gamma-ray flux reported here is broadly consistent with previous measurements by MAGIC \cite{MAGIC:2006pue, MAGIC:2008eqb} and VERITAS \cite{2017ICRC...35..712K}.

Benefiting from the broad energy coverage of LHAASO, we probe the orbital modulation across different energy bands. 
As shown in Fig.~\ref{fig3}, the phase-resolved flux measured by WCDA at 1.5--30.5 TeV extends over most phases and peaks near $\phi=0.6$. A likelihood-ratio test rejects the constant-flux hypothesis with a $p$-value of $0.0053$, corresponding to a significance of $\sim 2.6\sigma$ (one-sided hereafter), which indicates potential marginal variability.
In the 25--100 TeV range, the emission is more confined to $\phi = 0.3$--$0.6$, with a $p$-value of $4.65 \times 10^{-5}$ ($\sim 3.9\sigma$). Above 100 TeV, due to limited statistics, we split the data into two broad phase bins: 0.2--0.6 and 0.6--0.2. This yields a $2.4\sigma$ hint for modulation. Interestingly, the UHE photons tend to cluster in the $\phi = 0.6$--$0.2$ range, differing from the distribution observed at lower energies. However, the current significance remains too low to draw a firm conclusion.

\section{Discussion and conclusion}\label{sec12}
In this study, we report the first definite detection of gamma-ray emission up to the UHE range from the gamma-ray binary \lsi\ using LHAASO observations. The extension of the spectral energy distribution to energies nearly an order of magnitude higher than previous measurements (e.g., by MAGIC and VERITAS) not only challenges conventional acceleration models but also provides a unique opportunity to probe the balance between particle acceleration and energy losses in such systems. 

In the leptonic scenario, high-energy electrons upscatter stellar photons via inverse Compton (IC) scattering. However, in the UHE regime, the IC cross-section enters the Klein–Nishina limit (For a 100 TeV electron, $\gamma \varepsilon \gg m_e c^2$ with $\varepsilon = 5.5$ eV), significantly suppressing the emission efficiency. Furthermore, the emission spectrum is constrained by the maximum electron energy attainable in the binary system, which is governed by the balance between particle acceleration efficiency, cooling timescales, and the physical size of the acceleration region. Early studies \cite{2007MNRAS.380..320K, 2008MNRAS.383..467K} have shown that the maximum energy is sensitive to parameters such as the magnetic field strength, the efficiency of the acceleration mechanism, and the ambient photon density. 
Assuming an acceleration timescale $\tau_{\rm acc} = E/(\eta eBc) \approx 1.1(E/1{\rm TeV})(\eta/0.1)^{-1}(B/1{\rm G})^{-1}$ s with efficiency factor $\eta \approx 0.1$ and a magnetic field 
$B \approx 0.2$ G \cite{Zabalza:2010fw}, and a synchrotron cooling timescale $t_{\rm syn}=6 \pi m_e^2 c^3 /(\sigma_T B^2 E) \approx 400 (E/1 {\rm TeV})^{-1} (B/1{\rm G})^{-2}$ s, the maximum energy $E_{\rm max}$ can be obtained by equating $\tau_{\rm acc}$ to the synchrotron cooling timescale $t_{\rm syn}$, then $E_{\rm max} = \sqrt{6 \pi \eta e m_{\rm e}^2 c^4/(\sigma_{T}B)} \approx 40$ TeV.
Even if the acceleration efficiency is at its extreme maximum $\eta = 1$ (Bohm diffusion limit), the attainable energy is only 130 TeV.
The local magnetic field strength $B(\phi)$ varies substantially along the orbit, driven by factors such as changing orbital separation, shock compression, Be-disk crossings, and magnetic reconnection or turbulent amplification. Therefore, adopting an orbit-averaged $B \approx 0.2$ G \cite{Zabalza:2010fw} may significantly misestimate the local $E_{\rm max}$. Consequently, despite the detection of photons up to $>$200 TeV, a purely leptonic scenario cannot yet be ruled out.

Hadronic processes are expected to be most efficient near periastron, as suggested by the KM2A flux enhancement around this orbital phase. At these phases, the compact object encounters higher ambient densities, for instance through interaction with the circumstellar disk (see Supplementary Material for detailed estimations and other binary parameter settings discussed hereafter). Adopting the currently accepted orbital parameters of \lsi, the radial density profile of the stellar wind can be expressed as $n(r)= \dot{M}/(4\pi r^2 v_{w} m_{p})$, yielding a wind density at periastron of
$n(r_{\rm p}) \approx 1.9\times 10^8~{\rm cm}^{-3}$, while the characteristic density in the circumstellar disk at periastron
is $n_{\rm disk,peri} \sim 10^{11}~{\rm cm}^{-3}$.
The proton–proton cooling timescale is $t_{pp}=(n\kappa_{pp}\sigma_{pp}c)^{-1}$, with $\sigma_{pp} \approx 40$ mbarn and $\kappa_{pp} \approx 0.5$. In the absence of disk-crossing events, one obtains $t_{pp} \sim 100$ days, which is much longer than the orbital period and thus renders this scenario unlikely, whereas during disk crossing, $t_{pp} \sim 4.6$ hours. The interaction efficiency $f_{pp}$ also differs between disk-crossing ($f_{pp} \approx 0.06$) and non-disk-crossing ($f_{pp} \sim 10^{-4}$) episodes.
Since only about one third of the pion energy is channeled into $\pi^0 \to \gamma\gamma$, the gamma-ray luminosity is related to the proton power by $L_{\gamma} \approx f_{pp} L_{p}/3$. From the KM2A spectrum, the measured 25--100 TeV flux corresponds to a gamma-ray luminosity of $L_{\gamma} \approx 5\times 10^{31}~{\rm erg}~{\rm s}^{-1}$ for a distance of $d \approx 2$ kpc. The inferred proton power during disk-crossing phases is $L_{p} \approx 2.5 \times 10^{33}$ erg s$^{-1}$, which requires rather stringent conditions to channel a substantial fraction of the pulsar spin-down power into protons; nevertheless, the required energy budget remains compatible with the spin-down luminosity of a putative pulsar ($\dot{E} \approx 6\times 10^{33}~{\rm erg}~{\rm s}^{-1}$). In contrast, in the non-crossing scenario, the required proton power increases to $L_p \approx 1.5\times 10^{36}~{\rm erg}~{\rm s}^{-1}$, which significantly exceeds the available spin-down power.
While a purely hadronic origin challenges the available energy budget when considering the broadband emission, a mixed lepto-hadronic scenario naturally accommodates the constraints (see Supplementary Material for details).
This scenario is further supported by archival X-ray and TeV observations (see Fig.~\ref{mwl}), which exhibit correlated orbital modulation peaking at phase $\phi \sim 0.6$, and suggest a common leptonic origin for the emission at these phases.

Although the observed energy-dependent orbital modulation is only moderately significant, it nevertheless offers valuable insights into the physical conditions across the orbit. While the leptonic model naturally accounts for the WCDA emission via IC scattering, its extension to the UHE regime is challenged by Klein–Nishina suppression and synchrotron cooling limits described above. So we propose a secondary mixed scenario where the dominant particle population shifts with energy. In the 25–100 TeV band, the post-periastron emission peak sustained through inferior conjunction suggests a target-density-driven hadronic enhancement, while also marking a gamma-ray-transparent phase\cite{2006A&A...451....9D,2021heas.confE..54D}. This interpretation is reinforced by the simultaneous GeV orbital light curve (Fig.~\ref{mwl}), which also exhibits a prominent peak near periastron. The synchronization of GeV and multi-TeV signals strongly implies that the compact object encounters a substantially denser medium at this phase, e.g. Be-disk crossing event, which boosts both the inverse Compton emission and the $pp$ interaction rate. In contrast, the $>$100 TeV photons clustering toward apastron indicate a transition to an `acceleration-driven' regime. As described by Zabalza et al. (2013) \cite{2013A&A...551A..17Z}, orbital motion drives a Coriolis-terminated shock extending beyond the binary, leading to a weaker-field zone that suppresses synchrotron cooling. Although both leptons and hadrons can in principle be accelerated to UHE energies under these conditions, the Klein–Nishina suppression of IC scattering severely limits leptonic efficiency, leaving hadrons as the only viable candidate for the $>$100 TeV emission. This implies that \lsi\ operates as a Galactic PeVatron candidate. Regarding the orbital visibility, it is worth noting that the detected photons emerge in a phase window that avoids the strongest $\gamma\gamma$ opacity expected around superior conjunction \cite{2006A&A...451....9D, 2021heas.confE..54D}. Given the current low statistics in the $>$100 TeV band, these qualitative effects suffice to reconcile the apparent concentration of the handful of UHE photons at larger separations, while detailed quantitative modeling must await higher-significance data. Future observations with LHAASO, augmented by coordinated multi-wavelength campaigns, will provide a more comprehensive dataset. Specifically, phase-resolved spectroscopy at the highest energies, deeper searches for $>$100 TeV photons, and contemporaneous X-ray and radio monitoring (to track synchrotron cooling and shock activity) will be particularly valuable. Combined phase-resolved radiative-transfer and cascade modeling, constrained by these multi-band data, should be able to quantify the relative roles of inner/outer accelerators and test the scenario outlined above.

\vspace{\baselineskip}
\noindent{\bf Acknowledgements}
We would like to thank all staff members who work at the LHAASO site above 4400 meters above sea level year-round to maintain the detector and keep the water recycling system, electricity power supply, and other components of the experiment operating smoothly. We are grateful to the
Chengdu Management Committee of Tianfu New Area for the constant financial support for research
with LHAASO data. We appreciate the computing and data service support provided by the National
High Energy Physics Data Center for the data analysis in this paper. This research work is supported by the following grants:   National Key Research and Development Program of China under grants 2025YFE0202600, The National Natural Science Foundation
of China (NSFC) No.12522510, No.12393851, No.12393852, No.12393853, No.12393854, No.12205314, No.12105301, No.12305120, No.12261160362, No.12105294, No.U1931201, No.12375107, No.12273038, CAS Project for Young Scientists in Basic Research (No. YSBR-061), Youth Innovation Promotion Association
CAS (No.2022010, No.2023275) and in Thailand by the National Science and Technology Development Agency (NSTDA) and the National
Research Council of Thailand (NRCT) under the High-Potential Research Team Grant Program
(N42A650868).

\appendix

\renewcommand{\thefigure}{S\arabic{figure}}
\renewcommand{\thetable}{S\arabic{table}}
\setcounter{figure}{0}
\setcounter{table}{0}

\section{LHAASO detectors}
The Large High Altitude Air Shower Observatory (LHAASO) is situated on Haizi Mountain in Daocheng County, Sichuan Province of China, standing at an altitude of 4,410 meters above sea level. It is a composite extensive air shower (EAS) detector array consisting of three sub-arrays: the Kilometer Square Array (KM2A), the 78,000 m Water Cherenkov Detector Array (WCDA), and the Wide Field-of-view atmospheric Cherenkov Telescope Array (WFCTA). The gamma rays for a wide energy range spanning from hundreds of GeV to several PeV are detected by WCDA and KM2A.

The VHE data  ($0.8\ {\rm TeV}<E< 31\ {\rm TeV}$ ) were collected by the whole WCDA.  The data used in this study were obtained from the full-array operation of the WCDA from March 5, 2021, to July 31, 2024, with a total exposure time of 1124 days.
The zenith angle of the shower was required to be less than 50 degrees.
The $\gamma$/Proton separation parameter $\mathcal{P}$ (a parameter that characterizes the `clumpiness' of the air shower, termed PINCness, is detailed in \cite{Abeysekara:2017mjj})
is required to be less than 1.12, 1.02, 0.90, 0.88, 0.88, 0.84, and 0.84 for segments with  $N_{\rm{hit}}$ value of [30\text{--}60), [60\text{--}100), [100\text{--}200), [200\text{--}300), [300\text{--}500), [500\text{--}800) and  [800\text{--}2000].
For each $N_{\rm{hit}}$ segment, the sky map in celestial coordinates (right ascension and declination) was divided into a grid of $0.1^\circ \times  0.1^\circ$, which was filled with the number of the detected events according to their reconstructed arrival direction. The `direct integration method' is adopted to estimate the number of CR background events in each pixel.

The data with energy above 25 TeV collected by KM2A are used in this analysis. KM2A comprises 5,216 electromagnetic particle detectors (EDs) and 1,188 underground muon detectors (MDs) spread over an area of approximately 1.3 km$^2$ \cite{hhh2018}. The KM2A was incrementally developed and began operations, with half of the array starting scientific observations at the end of 2019, three-quarters operational by December 2020, and full operations officially launched on July 31, 2024, with a total exposure time of 1570.42 days. Only showers with reconstructed zenith angle less than 50$^\circ$ are used.  The ratio of measured muons to electrons is utilized to discriminate primary $\gamma$-rays from cosmic nuclei. Energies are estimated based on particle density at a radius of 50 m (denoted as $\rho_{50}$) across various zenith angles. The energy resolution is better than 20\% for energies above 100 TeV within the zenith angle range of $0^\circ$--$35^\circ$. The data sets are categorized into five groups per decade based on reconstructed energy. The sky map is represented in celestial coordinates, binned into a grid of $0.1^{\circ} \times 0.1^{\circ}$ pixels, with each pixel filled with the number of detected events corresponding to their reconstructed arrival direction. The `direct integration method' is employed to estimate the cosmic ray background. To optimize array performance, several event selection criteria are implemented. The data selection criteria published in the KM2A performance paper\cite{LHAASO:2021ozi} are used in this study.

\begin{table}[!htp]
\caption{Energy bin definitions for WCDA (upper) and KM2A (lower) and the characteristics of \lsi\ at each energy bin. Note that the definitions for WCDA and KM2A are different: WCDA data is binned depending on the number of triggered detector units($N_{\rm{hit}}$), and KM2A data is divided by the reconstructed energy. }
\label{table:Flux point}
\begin{center}
\begin{tabular*}{\textwidth}{@{\extracolsep{\fill} } cccc}
\hline
\hline
\textbf{N$_{\rm{hit}}$} & \textbf{Median energy (TeV)} &\textbf{Differential Flux ($\textrm    {TeV}^{-1}\textrm{cm}^{-2}\textrm{s}^{-1}$)} & \textbf{TS}\\
\hline
         30-60    & 0.83  & $<3.95 \times 10^{-12}$ (95\% UL) & 0.7\\
         60-100   & 1.41  & $(5.23 \pm 2.21) \times 10^{-13}$ & 5.6\\
         100-200  & 2.49  & $(1.42 \pm 0.27) \times 10^{-13}$ & 28.5\\
         200-300  & 4.29  & $(3.42 \pm 0.66) \times 10^{-14}$ & 28.9\\
         300-500  & 6.91  & $(6.45 \pm 1.61) \times 10^{-15}$ & 18.3\\
         500-800  & 14.72 & $(5.12 \pm 2.26) \times 10^{-16}$ & 6.7\\
         800-2000 & 31.33 & $(1.12 \pm 0.37) \times 10^{-16}$ & 14.4\\
\hline
\hline
\textbf{log$_{10}$($E_{\mathrm{rec}})$} & \textbf{Median energy (TeV)} &\textbf{Differential Flux ($\textrm    {TeV}^{-1}\textrm{cm}^{-2}\textrm{s}^{-1}$)} & \textbf{TS}\\
\hline
         1.4-1.6  & 28.18  & $(1.17 \pm 0.62)\times 10^{-16}$  & 4.0\\
         1.6-1.8  & 44.67  & $(3.45 \pm 0.96) \times 10^{-17}$ & 17.5\\
         1.8-2.0  & 70.79  & $(7.47 \pm 2.26) \times 10^{-18}$ & 13.8\\
         2.0-2.2  & 112.20  & $(1.22 \pm 0.50) \times 10^{-18}$ & 8.9\\
         2.2-2.4  & 199.53  & $(2.10 \pm 1.07) \times 10^{-19}$ & 6.6\\
         2.4-3.0  & 501.19 & $<2.35\times 10^{-20}$(95\% UL) & 1.9\\
\hline
\hline
\end{tabular*}\\
\end{center}
\end{table}

\section{Data analysis}
\subsection{Likelihood fitting}
\begin{figure}[h]
\centering
\includegraphics[width=0.45\textwidth]{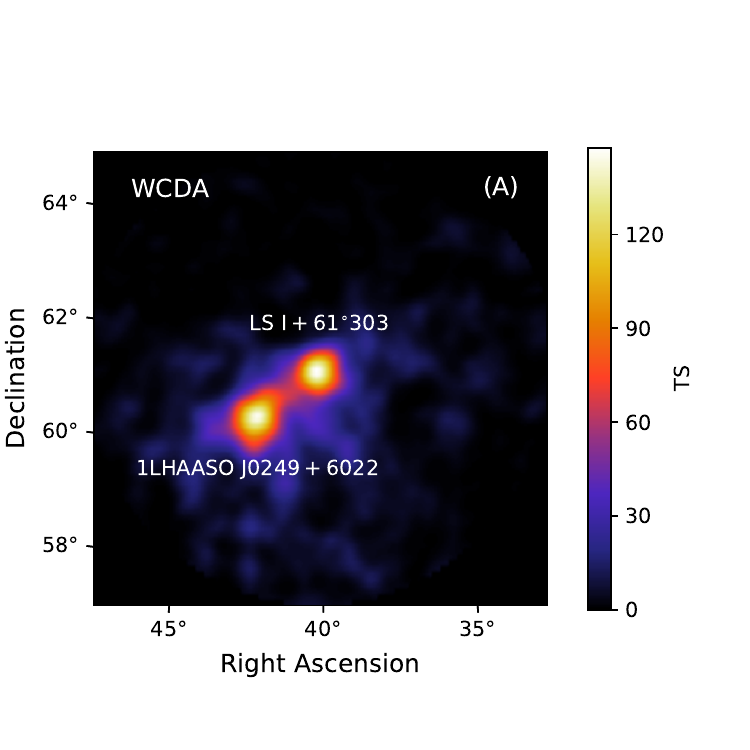}
\includegraphics[width=0.45\textwidth]{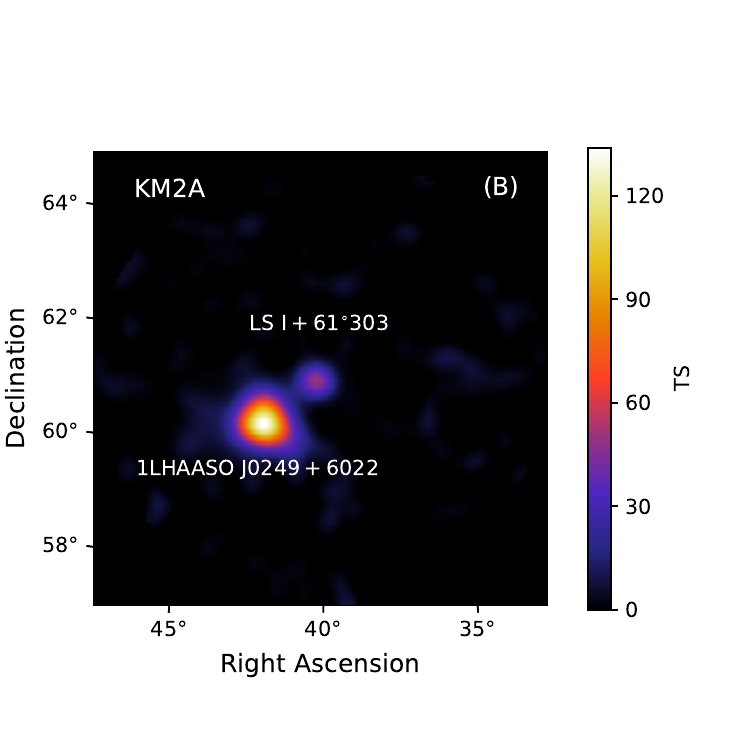}
\caption{Test statistic (TS) maps of the region around \lsi\ observed by WCDA {\bf (A)} and KM2A {\bf (B)}. The color scale represents the TS value of the excess. The target source \lsi\ is clearly visible in the center. The neighboring source 1LHAASO J0249+6022, located $\sim 1.9^\circ$ away, is likely a PWN or pulsar halo associated with PSR J0248+6021. The large-scale extended features are attributed to the Galactic diffuse emission (GDE).
}
\label{fig4}
\end{figure}

Considering the influence from Galactic diffuse gamma-ray emission (GDE) and 1LHAASO J0249+6022 close to \lsi, we define a 4$^\circ$-radius circular ROI around \lsi, centered at R.A. = 40.1$^\circ$, Decl. = 61.2$^\circ$ (see Fig. \ref{fig4}). The data analysis is based on a 3-dimensional likelihood fitting method, where the position and spectrum for each source in the Region of Interest (ROI) are fitted simultaneously. The morphology of the source is described by a 2D Gaussian function. Two kinds of spectrum forms are tested in this analysis, which are the simple power-law function and the exponential cut-off function:
\begin{equation}
dN/dE=F\times(E/E_{0})^{-\alpha} 
\label{func1}
\end{equation} 

\begin{equation}
dN/dE=F\times(E/E_{0})^{-\alpha}\times e^{-E/E_{c}} 
\label{func2}
\end{equation} 

Where $F$ is the differential flux normalization at $E_{0}$. $E_{0}$ is the reference energy, which is chosen to be 50 TeV for KM2A and 3 TeV for WCDA separately. $\alpha$ is the spectral index, and E$_{c}$ is the cut-off energy.

The observed number of events in each pixel is assumed to follow a Poisson distribution, with the mean value being the sum of cosmic-ray background counts and expected gamma-ray counts from sources and the GDE. The expected gamma-ray counts are calculated by folding the source model through the instrument response (obtained from detector simulations). 
The likelihood value $\mathcal{L}$ is defined as the product of the probabilities of each pixel.
It should be explicitly noted that in this 3D likelihood maximization, the cosmic-ray background is fixed to the value derived from the direct integration method. This approach corresponds to the Cash statistic \citep{1979ApJ...228..939C}. Ignoring the statistical uncertainty of the background in the fit is practically justified because the direct integration method leverages a massive off-source data pool (resulting in an effective ON/OFF exposure ratio $\alpha \ll 1$), rendering the statistical error on the background estimation negligible compared to the fluctuations of source counts.
However, in scenarios where the cosmic-ray background heavily dominates, its associated uncertainty becomes critical and should be explicitly evaluated (e.g., \cite{LHAASO_2023dge}). Conversely, in the $>100$ TeV regime, contributions from the GDE and 1LHAASO J0249+6022 become negligible. This spatial simplicity naturally allows us to employ the standard Li \& Ma formula \citep{LiMa}, which rigorously incorporates the background statistical uncertainty but is otherwise difficult to apply in complex, multi-component regions.

The likelihood ratio test is used to perform the hypothesis testing. The Test Statistic (TS) is defined as:
\begin{equation}
    \rm TS=-2(ln\mathcal{L}_{0}-ln\mathcal{L}_{1})
\end{equation}
where $\mathcal{L}_{0}$ and $\mathcal{L}_{1}$ denote the maximum likelihood values obtained under the null and alternative hypotheses, respectively. The TS is used to evaluate the goodness of fit between the two hypotheses and is assumed to follow a $\chi^2$ distribution (according to Wilks' theorem). To determine the number of sources in the ROI, an iterative likelihood ratio test is performed. In each step, the null hypothesis represents the current model, while the alternative hypothesis adds a new point source to the model. The significance of the new source is calculated using $\Delta$TS associated with 5 degrees of freedom (corresponding to the number of free parameters). If $\Delta$TS exceeds 25 (corresponding to a significance of ~3.6$\sigma$), this new source is added to the model for the next iteration. This process repeats until no new candidate meets the criterion. Once the source list is determined, the significance of each source is re-evaluated by fixing the parameters of GDE and other sources. The flux points are obtained by fitting the normalization independently at each energy bin. Finally, a TS map is derived using the same likelihood ratio test method applied to each pixel. Since the large-field maps contain multiple sources, we employed fixed `typical' indices (-2.6 for WCDA and -3.0 for KM2A) as a standardized baseline for these maps.

\noindent

\subsection{Results}
A total of two sources are resolved within the ROI, the other being 1LHAASO J0249+6022, which is likely a pulsar halo or pulsar wind nebula (PWN) associated with PSR J0248+6021. The Test Statistic ($\Delta {\rm TS}$) for \lsi\ is 51.8 in the KM2A energy range, corresponding to a significance of 6.2$\sigma$. We also fit the data with an exponential cutoff power-law model to test for potential curvature in the spectral energy distribution (SED) of \lsi. Compared to the pure power-law fit, the TS value increases by only 2 with the addition of one extra parameter. We therefore conclude that the pure power-law model is statistically favored, and no significant cutoff is observed in the SED above 25 TeV. Similarly, the WCDA SED is well described by a power law. Details of the median energy, flux, and TS for each energy bin are summarized in the Table.~\ref{table:Flux point}.

\begin{figure}[htbp]
\centering
\includegraphics[width=0.45\textwidth]{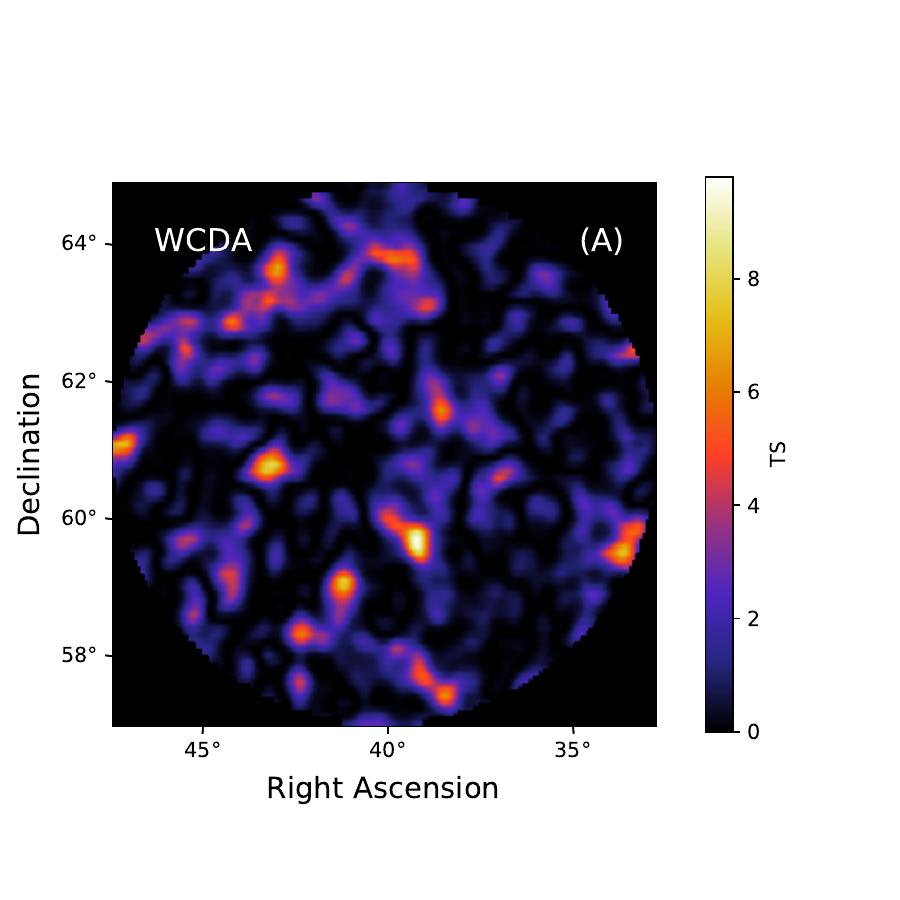}
\includegraphics[width=0.45\textwidth]{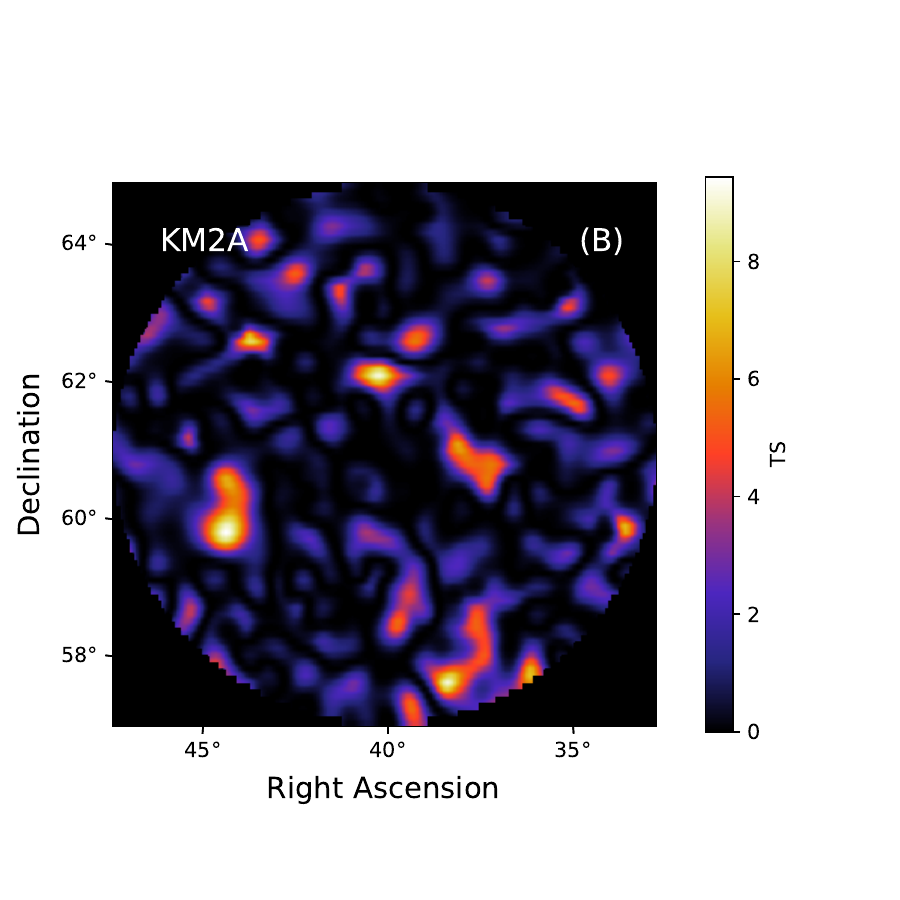}
\caption{The 2D residual test statistic (TS) maps from WCDA {\bf (A)} and KM2A {\bf (B)} are shown after subtracting the best-fit models for \lsi, 1LHAASO J0249+6022, and the Galactic diffuse emission. Each pixel represents the TS value derived from a likelihood ratio test. No residual excess with $\sqrt{\text{TS}} > 3$ is observed in either map.}
\label{fig5}
\end{figure}

\begin{figure}[htbp]
\centering
\includegraphics[width=0.4\textwidth]{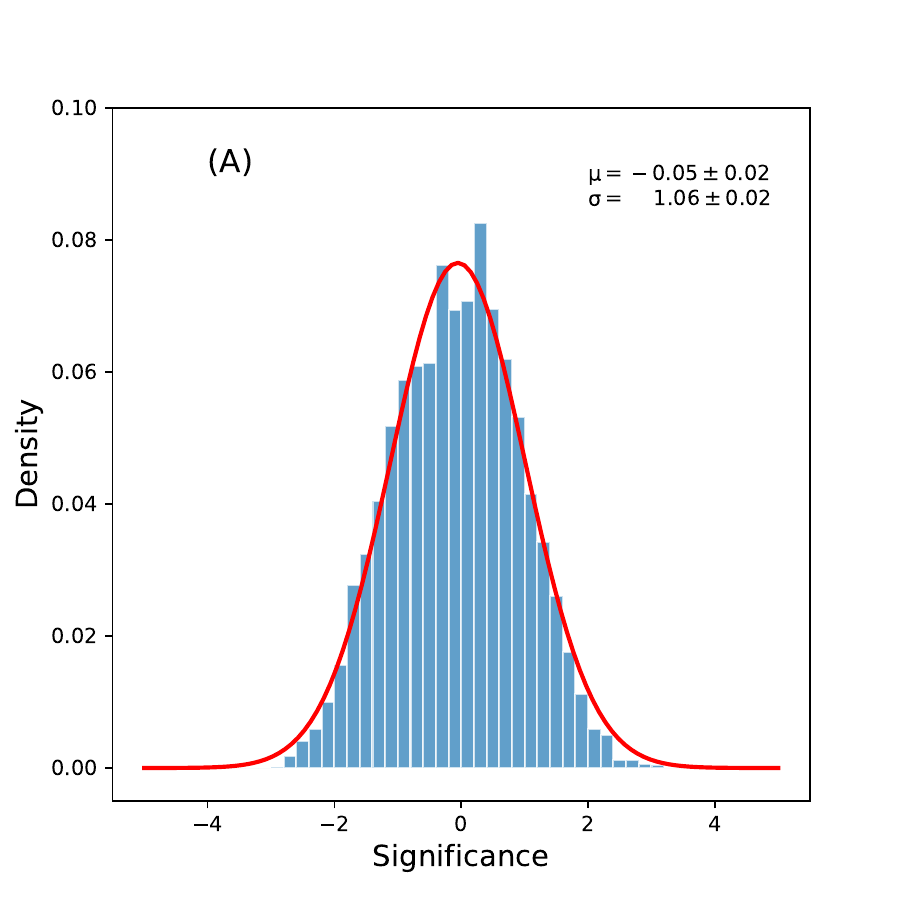}
\includegraphics[width=0.4\textwidth]{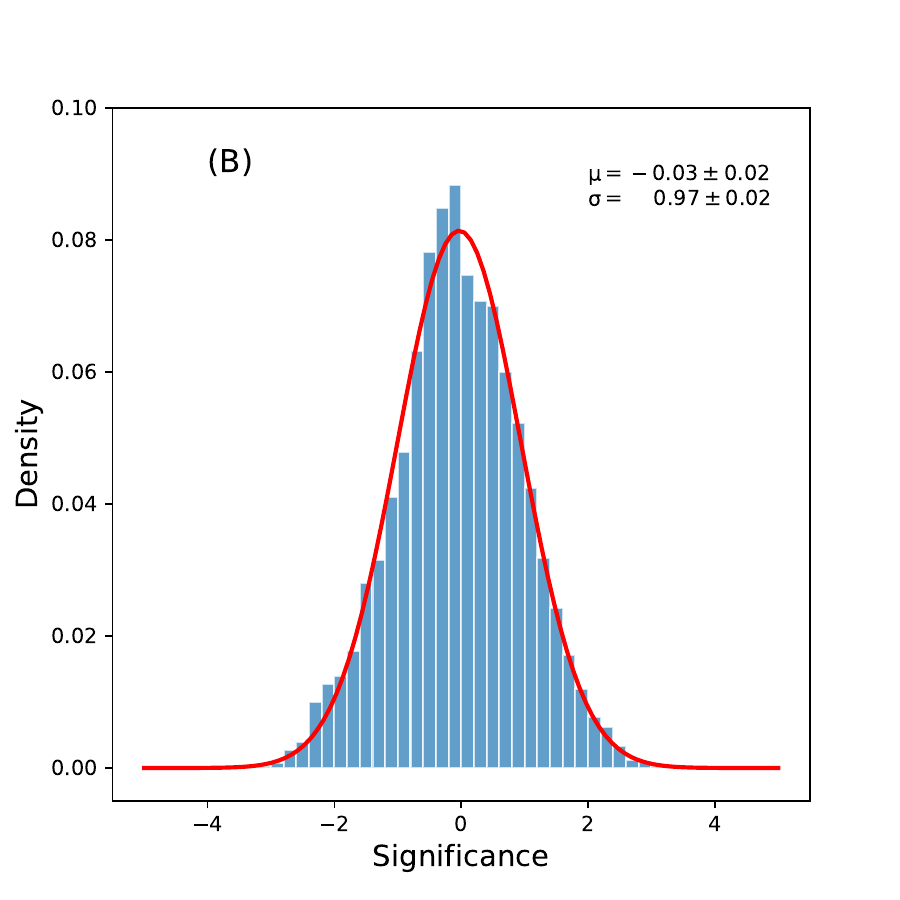}
\caption{The 1D histograms show the distribution of $\sqrt{\text{TS}}$ values extracted from the residual maps (Fig.~\ref{fig5}) for WCDA {\bf (A)} and KM2A {\bf (B)}. The red solid lines represent the best-fit results using a Gaussian function. Both distributions are consistent with a standard Gaussian (mean $\approx 0$, $\sigma \approx 1$), indicating that the residuals are dominated by statistical fluctuations and the background is properly modeled.}
\label{fig6}
\end{figure}

Fig.~\ref{fig5} shows the residual map after subtracting all modeled sources within the ROI, in which no significant excess is visible. The distribution of TS values from the map is converted into statistical significance according to Wilks' theorem and is well-fitted by a Gaussian distribution, as shown in Fig.~\ref{fig6}. For WCDA, the fit yields a mean of $-0.05 \pm 0.02$ and a standard deviation of $1.06 \pm 0.02$. For KM2A, we obtain a mean of $-0.03 \pm 0.02$ and a standard deviation of $0.97 \pm 0.02$.

To evaluate the systematic pointing uncertainty at the declination of \lsi, we analyzed the position of the Crab Nebula using events with zenith angles $\theta > 30^\circ$, weighting the zenith distribution to match the transit path of \lsi. For KM2A, the maximum observed deviations were $0.036^\circ \pm 0.015^\circ$ in R.A. and $0.028^\circ \pm 0.016^\circ$ in Decl. Although these values are likely dominated by statistical limitations, we conservatively adopt a systematic pointing error of $0.05^\circ$ for KM2A. For WCDA, the same procedure yielded a positional deviation of less than $0.01^\circ$. We further cross-verified this accuracy using multiple extragalactic sources, including Mrk 421, Mrk 501, and 1ES 1959+650 (located at a declination similar to \lsi); these cross-checks confirm that the WCDA pointing error is better than $0.03^\circ$.

\begin{figure}[htbp]
\centering
\includegraphics[width=0.6\textwidth]{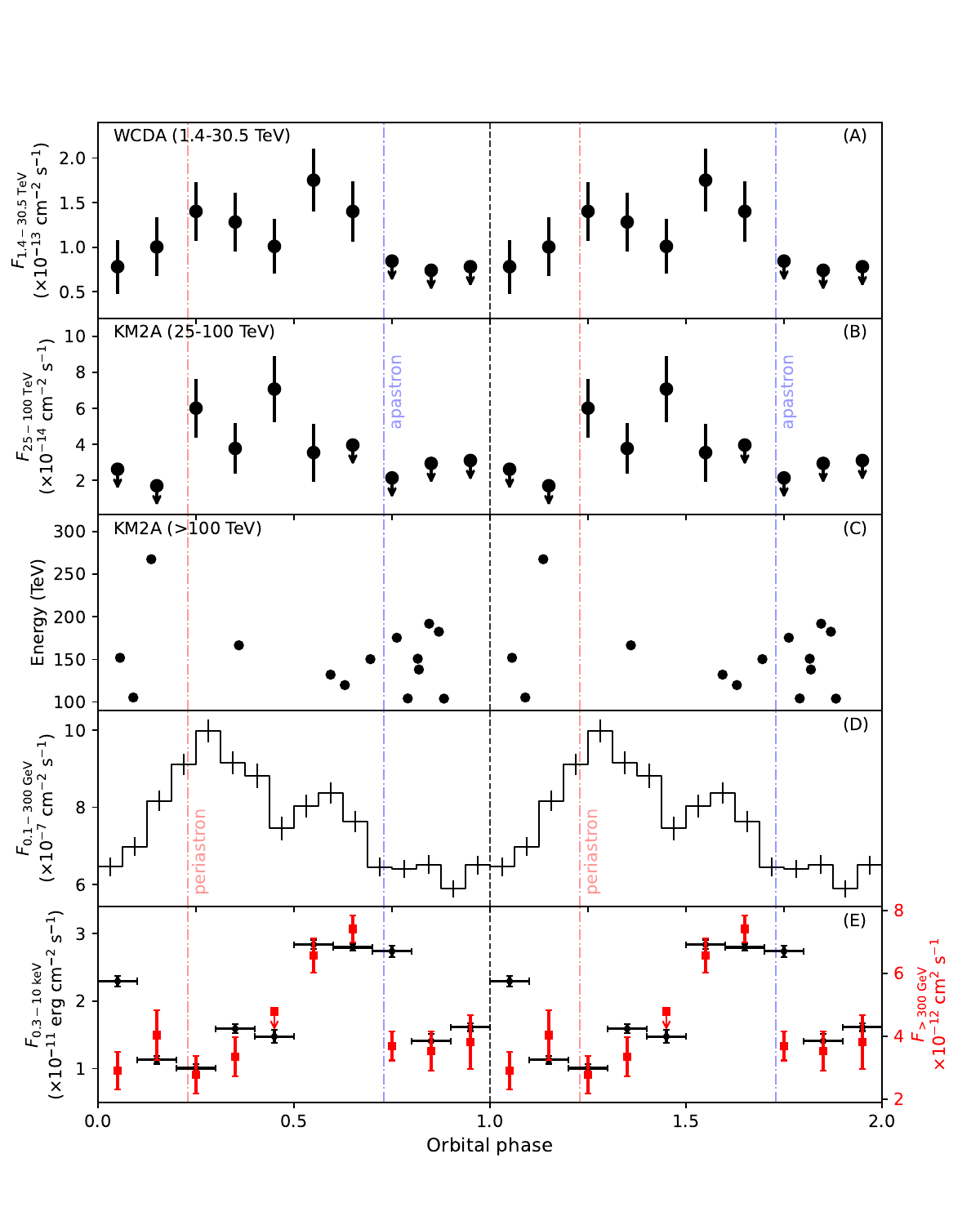}
\caption{Orbital modulations at multiple wavelengths. ({\bf A,B,C}) are the same with Fig.~\ref{fig3}. ({\bf D}) shows the light curve derived from simultaneous Fermi-LAT observations. ({\bf E}) Although the Swift-XRT \cite{2007A&A...474..575E} and VERITAS \cite{2017ICRC...35..712K} data were not taken contemporaneously with this work, their peak fluxes both occur near $\phi \sim 0.6$.} 
\label{mwl}
\end{figure}

\subsection{Fermi-LAT data analysis}
To obtain the Fermi orbital light curve contemporaneous with LHAASO observations (since the source exhibits different orbital modulation patterns at different superorbital phases), we analyzed approximately three years of data (MJD 59415–60523). The PASS 8 SOURCE class events were extracted from a $20^{\circ} \times 20^{\circ}$ region of interest (ROI) centered on \lsi, with an energy range of 0.1–300 GeV. The analysis was performed using Fermitools v2.2.0, adopting the instrument response function P8R3\_SOURCE\_V3, and the background model was constructed based on the 4FGL-DR4 catalog \cite{Fermi-LAT:2019yla}, along with the Galactic diffuse component (gll\_iem\_v07.fits) and the isotropic extragalactic component (iso\_P8R3\_SOURCE\_V3\_v1.txt). The globally fitted model was subsequently used as the baseline for the orbital flux reconstruction. In practice, the orbital period was evenly divided into 16 phase bins, after which we performed phase-resolved likelihood analyses. Specifically, events were re-filtered within each phase interval following period folding, and the likelihood fit was independently repeated to derive the flux in each phase bin. Given the smoothing effect introduced by folding over the orbital period of \lsi, we assumed that background sources do not vary with this orbital phase. Therefore, during the phase-resolved fitting, only the spectral parameters of \lsi\ and normalizations of the diffuse components were left free, while all other background sources were fixed to their global best-fit values obtained above.

\section{Binary parameter setting}
In the discussion, we adopt parameter values established in previous studies as the baseline. The temperature of the Be star is taken to be 22.5 kK \cite{1995A&A...298..151M}, corresponding to a characteristic blackbody photon energy of $E \sim k_{\rm B} T = 1.94$ eV. The peak of the SED occurs at $E_{\rm peak} \approx 5.5$ eV, which lies in the ultraviolet band. According to the model proposed by Millar \& Marlborough (1998) \cite{1998ApJ...494..715M}, the temperature in the inner region of the circumstellar disk is approximately 60\% of the stellar temperature, yielding $T_{\rm disk,0} \approx 0.6T_{*} \approx 13500$ K. For viscous-disk, the radial temperature profile can be estimated using $T(r) = T_{\rm disk,0}(r/R_*)^{-3/4}$ \cite{1974MNRAS.168..603L}. To determine the specific environment encountered by the compact object, we must first constrain the orbital distance. Since the orbital eccentricity of the binary is known ($e \sim 0.54$ \cite{2009ApJ...698..514A}), the binary separation at periastron can be calculated as $r_{\rm peri}=a (1-e)$, where the semi-major axis $a=0.42$ AU is obtained from Kepler’s law. This gives $r_{\rm peri} \approx 0.19$ AU. Under additional parameter assumptions ($R_*=10R_{\odot}$ and a disk radius $R_d=7R_* \approx 0.33$ AU \cite{Zabalza:2010fw}), it becomes evident that the compact object may undergo a disk-crossing event when passing through periastron.

To further estimate the matter density variations encountered by the compact object during the potential periastron disk-crossing event, here we adopt canonical values for the mass-loss rate $\dot{M} \sim 5\times 10^{-8} M_{\odot}~{\rm yr}^{-1}$ and velocity $v_w=10^{8}~{\rm cm}~{\rm s}^{-1}$. At periastron, this yields $n(r_{\rm p}) \approx 1.9\times 10^8~{\rm cm}^{-3}$. Within the Be-star circumstellar disk, the particle density is expected to exceed that of the stellar wind by approximately 2–4 orders of magnitude, depending on the disk structure and radial distance (e.g., \cite{2013A&ARv..21...69R}). For the purposes of our estimates, we adopt a fiducial disk density of $n_{\rm disk,peri} \sim 10^{11}~{\rm cm}^{-3}$ at periastron.
In the $pp$ scenario, the interaction efficiency can be estimated as $f_{pp} = t_{\rm esc}/(t_{pp}+t_{\rm esc})$, where the escape/advection timescale is aproximated by $t_{\rm esc} \sim l/v$. For a compact interaction region with a characteristic size $l \sim 10^{12}$ cm and a flow velocity $v \sim 10^9~{\rm cm}~{\rm s}^{-1}$, this yields $t_{\rm esc} \sim 10^3$ s. As a result, we obtain $f_{pp} \approx 0.06$ during disk-crossing episodes, while it decreases to $f_{pp} \approx 10^{-4}$ outside these phases.

The KM2A energy flux in the 25--100 TeV band is measured to be $F_{\gamma,{\rm 25-100 TeV}} = 1.05\times 10^{-13}~{\rm erg}~{\rm cm}^{-2}~{\rm s}^{-1}$. Assuming a source distance of $d \approx 2.0$ kpc, the corresponding gamma-ray luminosity is $L_{\gamma} = 4\pi d^2 F_{\gamma} \approx 5\times 10^{31}~{\rm erg}~{\rm s}^{-1}$. The proton power required to sustain this gamma-ray luminosity is given by $L_{p} \approx 3L_{\gamma}/f_{pp}$, where $f_{pp}$ denotes the effective $pp$ interaction efficiency. To assess whether such a proton power can be supplied by a pulsar, we estimate the spin-down luminosity assuming a neutron star with period $P \simeq 270$ ms \cite{Weng:2022zlg}, moment of inertia $I \approx 10^{45}~{\rm g}~{\rm cm}^2$, and period derivative $\dot{P}=3\times 10^{-15}~{\rm s}~{\rm s}^{-1}$ \cite{2005AJ....129.1993M}. The spin-down power is then $\dot{E} = 4\pi^2 I \dot{P}/P^3 \approx 6\times 10^{33}~{\rm erg}~{\rm s}^{-1}$.
However, integrating the measured soft power-law spectrum ($\Gamma \approx 3.0$) above 1 TeV yields a bolometric gamma-ray luminosity of $L_{\gamma, >1{\rm TeV}} \approx 1.7 \times 10^{33}~{\rm erg}~{\rm s}^{-1}$. In a purely hadronic scenario with an interaction efficiency of $f_{pp} \approx 0.06$, the required proton power would be $L_p \approx 3 L_{\gamma} / f_{pp} \approx 8.5 \times 10^{34}~{\rm erg}~{\rm s}^{-1}$, which exceeds the available pulsar spin-down power ($\dot{E} \approx 6 \times 10^{33}~{\rm erg}~{\rm s}^{-1}$) by over an order of magnitude. This energy budget crisis is naturally resolved in a mixed lepto-hadronic scenario. Since electrons are in the fast-cooling regime ($f_{\rm IC} \approx 1$; $t_{IC} \ll t_{\rm esc}$) due to the intense stellar radiation field, they can efficiently power the bulk emission below tens of TeV (we take 25 TeV here, which requires $L_e \approx 1.7 \times 10^{33}~{\rm erg}~{\rm s}^{-1}$). Protons are thus only required to sustain the high-energy tail ($>25$ TeV, $L_{\gamma} \approx 5 \times 10^{31}~{\rm erg}~{\rm s}^{-1}$); with $f_{pp} \approx 0.06$, this demands $L_p \approx 2.5 \times 10^{33}~{\rm erg}~{\rm s}^{-1}$. The total injection power $L_{\rm inj} \approx 4.2 \times 10^{33}~{\rm erg}~{\rm s}^{-1}$ fits comfortably within $\dot{E}$, favoring a mixed origin where protons dominate only the highest energies.
The temporal correlation between X-ray and VHE gamma-ray fluxes peaking at $\phi \sim 0.6$ (Fig.~\ref{mwl}) implies a common leptonic origin, where the same electron population produces synchrotron X-rays and Inverse Compton gamma-rays. Identifying this distinct leptonic component confirms the validity of the mixed scenario and significantly alleviates the energy budget crisis that arises if the entire broadband emission is attributed solely to hadronic processes.


\bibliographystyle{apsrev4-2}
\bibliography{sn-bibliography}

\end{document}